\documentclass[11pt,twoside]{article}

\usepackage{asp2006}

\def\ha{H$\alpha$}

\def\mi{$\mu$m}
\def\d4n{D$_n$(4000)}
\def\ltir{$L_{TIR}$}

\newcommand\newion[2]{#1$\;${\scshape{#2}}}

\def\neii{[\newion{Ne}{ii}]12.8$\mu$m}

\def\siii{[\newion{S}{iii}]18.7$\mu$m}
\def\oiv{[\newion{O}{iv}]25.9$\mu$m}
\def\o3hb{[\newion{O}{iii}]$\lambda 5007$/H$\beta$}
\def\n2ha{[\newion{N}{ii}]$\lambda 6583$/H$\alpha$}

\begin{document}
\title{Mid-Infrared Spectral Indicators of Star-Formation and AGN Activity in Normal Galaxies}   
\author{Marie Treyer} 
\affil{California Institute of Technology, MC 278-17, 1200 E. California Boulevard, Pasadena, CA 91125, USA}
\author{Ben Johnson}
\affil{Institute of Astronomy, University of Cambridge, Madingley Road, Cambridge CB3 0HA, UK}
\author{David Schiminovich and Matt O'Dowd}
\affil{Astronomy Department, Columbia University, 550 W. 120 St., New York, NY 10027,USA}

\begin{abstract} 
We investigate the use of mid-infrared PAH bands, 
continuum and emission lines as probes of star-formation and AGN activity 
in a sample of 100 `normal' and local ($z \sim 0.1$) galaxies. 
The MIR spectra were obtained with the {\it Spitzer} IRS 
as part of the Spitzer-SDSS-GALEX Spectroscopic Survey (SSGSS) which includes multi-wavelength
photometry from the UV to the FIR and optical spectroscopy. The spectra were decomposed
using PAHFIT (Smith et al. 2007), which we find to 
yield PAH equivalent widths (EW) up to $\sim 30$ times larger than the commonly used spline methods. 
Based on correlations between PAH, continuum and emission line properties and optically derived 
physical properties (gas phase metallicity, radiation field hardness), 
we revisit the diagnostic diagram relating PAH EWs and \neii/\oiv\ and find
it more efficient as distinguishing weak AGNs from star-forming galaxies than
when spline decompositions are used.  
The luminosity of individual MIR component (PAH, continuum, Ne and H$_2$ lines)
are found to be tightly correlated to the total IR luminosity and can be used to estimate 
dust attenuation in the UV and in \ha\ lines based on energy balance arguments.
\end{abstract}
\section{Goals}
We aim at determining the main source of ionizing radiation and
star-formation rate of normal galaxies from MIR spectroscopy.

\section{The SSGSS Sample}
The Spitzer-SDSS-GALEX Spectroscopic Survey is an IRS survey of 100 local galaxies in the Lockman Hole. 
The data include GALEX FUV photometry, SDSS optical imaging and spectroscopy, Spitzer IRAC and MIPS photometry. 
The sample has a surface brightness limit of 0.75 MJy sr$^{-1}$ at 5.8\mi\ and a flux limit of 1.5mJy at 24\mi.
It was selected to cover the range of physical properties of `normal' galaxies (e.g.
$9.3\le {\rm log}(M/M_\odot) \le 11.3$, $8.7\le {\rm log(O/H)}+12\le 9.2$, $0.4<A_{{\rm H}\alpha}<2.3$).
The redshifts span $0.03<z<0.21$ with a mean of 0.1 similar to that of the full SDSS spectroscopic sample. 
Galaxies are classified as star-forming (black dots), composite 
(pink stars) or AGN (red triangles) according to the boundaries 
of \citet{Kewley_etal2001} and \citet{Kauffmann_etal2003} on the \n2ha\ versus \o3hb\  ``BPT'' diagram 
\citep{BPT1981}. First results were reported by \citet{ODowd_etal2009}.

\begin{figure}[t]
\plottwo{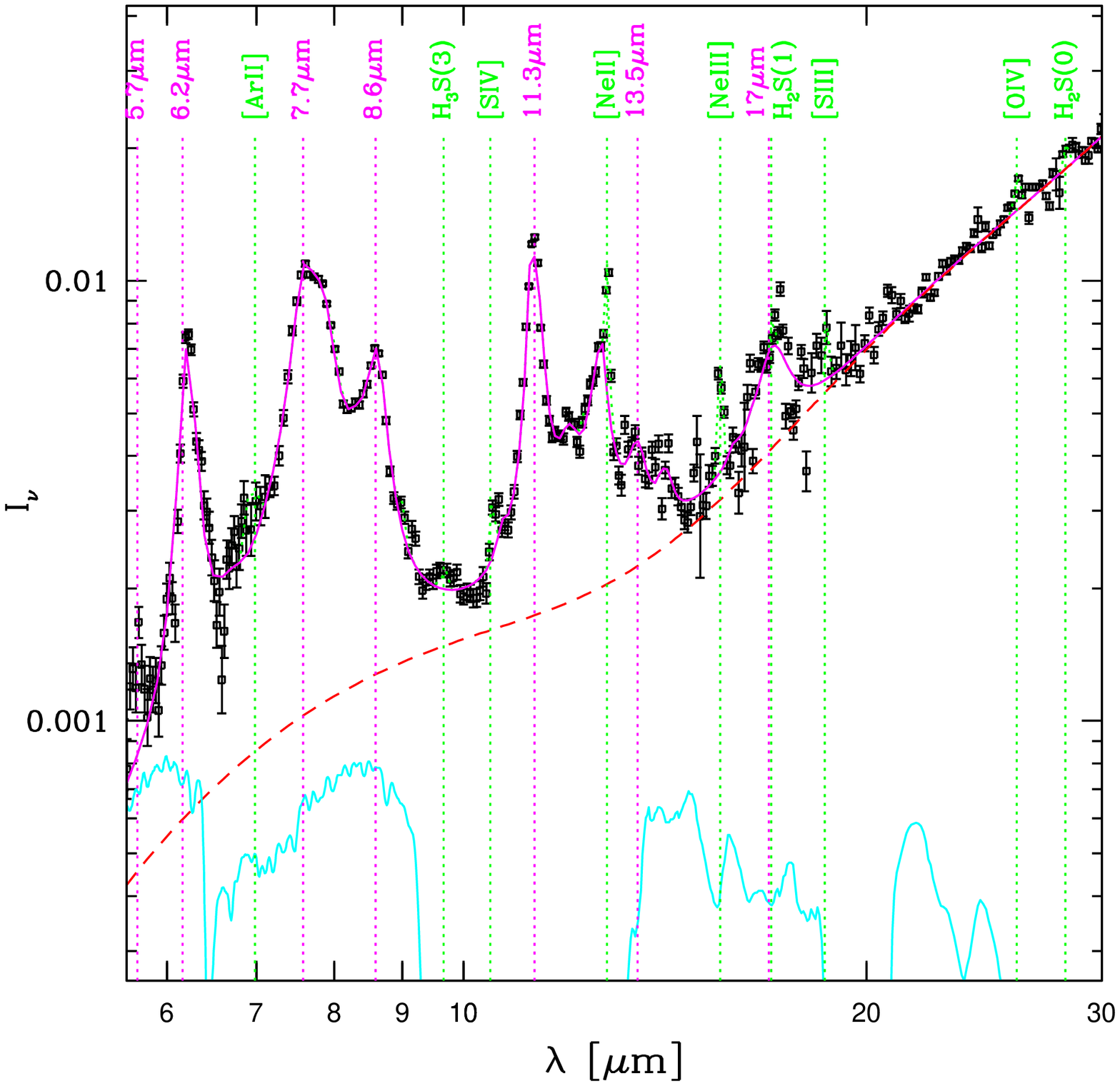}{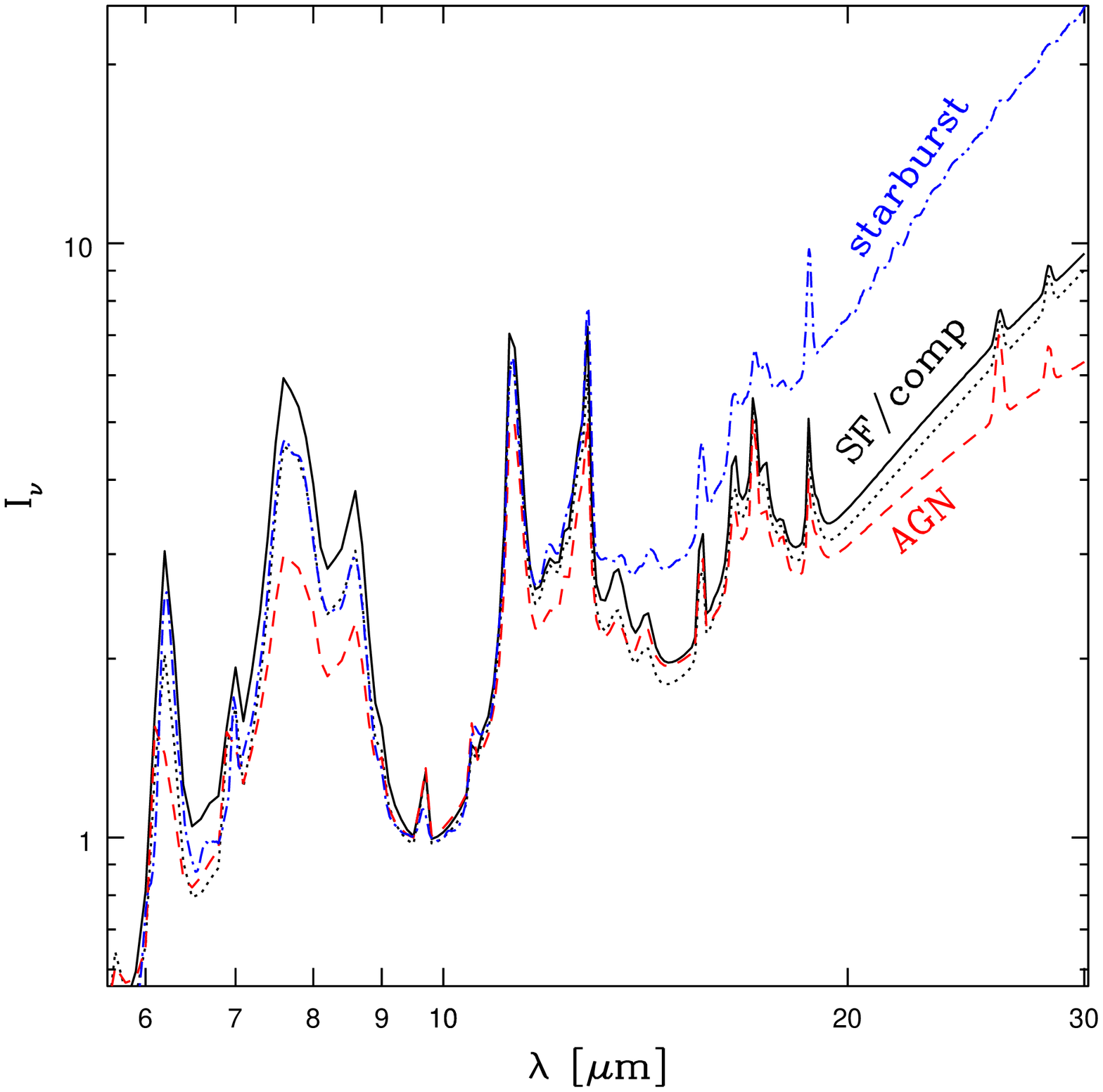}
\caption{{\it Left:} Example spectrum (a SF galaxy) with PAHFIT decomposition: continuum (dashed red line),
PAH features (solid purple line) and emission lines (dotted green line).
The lower curves show - from left to right - the filter responses of the IRAC bands at 6 and 8\mi, 
of the IRS blue Peak Up band at 16\mi\ and of the MIPS band at 24\mi\ bands.  
{\it Right:} The mean spectra of SF galaxies (solid line), composite galaxies (dotted line) and AGNs (dashed line). 
The dot-dashed line is the average starburst spectrum of \cite{Brandl_etal2006}. The spectra are normalized at 10\mi.
\label{fig:fig1}
}
\end{figure}

\section{Spectral Decomposition}

We used PAHFIT \citep{Smith_etal2007} to decompose the spectra into a sum of dust attenuated thermal dust continuum, PAH features 
and emission lines. The left panel of Fig. 1 shows an example decomposition for a typical SF galaxy. The right panel
shows the mean spectra of SF galaxies, composite galaxies and AGNs along with the
average starburst spectrum of \citet{Brandl_etal2006}. 
The transition from starburst to SF galaxy to AGN is marked by a declining continuum slope, decreased \neii\ and \siii\ 
emission and enhanced \oiv\ emission. 
The AGN spectrum, and to a lesser extent the starburst spectrum, show weaker PAH emission at low wavelength than 
the SF spectrum, an effect attributed to the destruction of PAHs in intense far-UV radiation fields.

\begin{figure}[t]
\plottwo{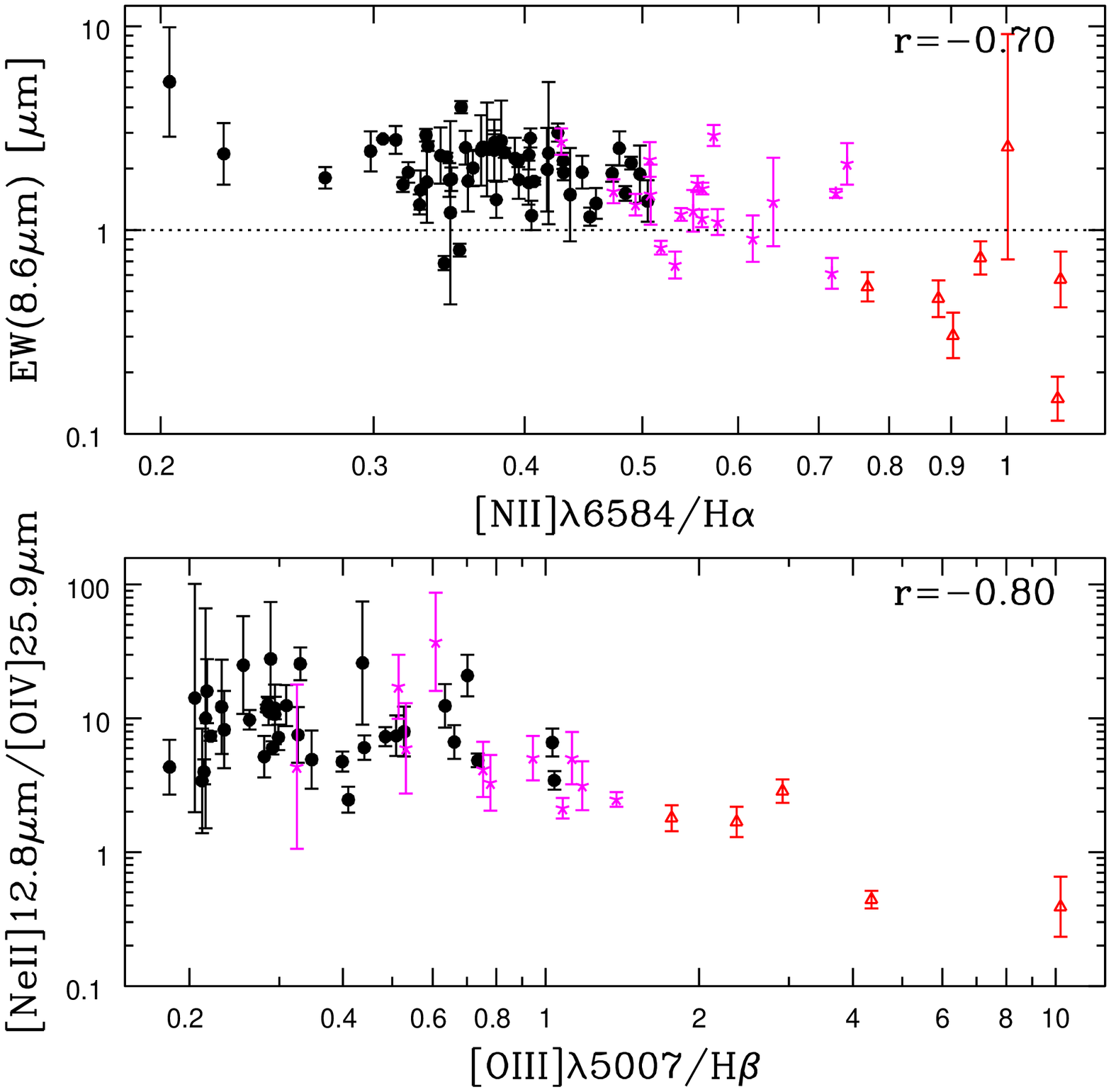}{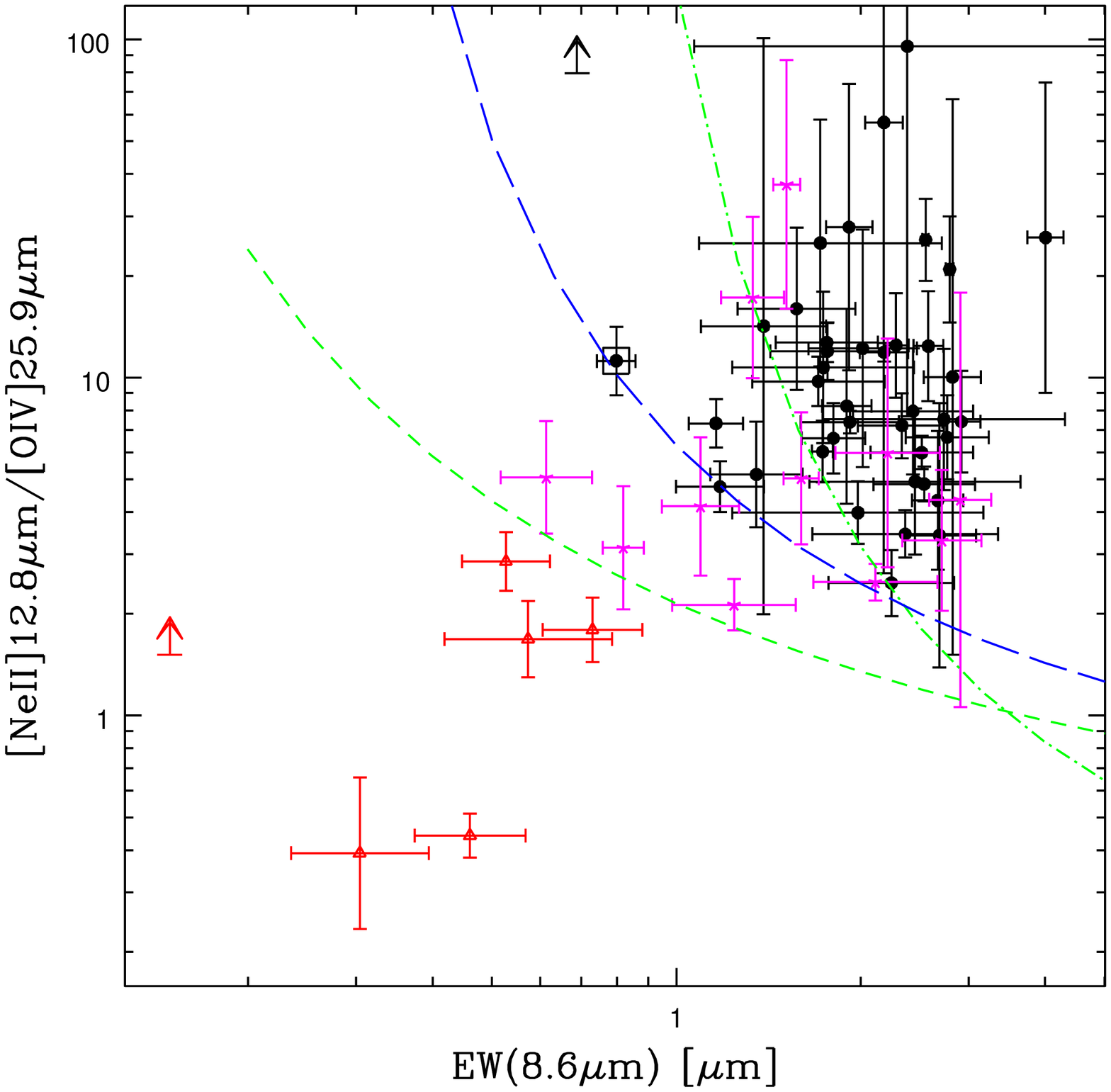}
\caption{{\it Top Left:} 8.6\mi\ PAH EW versus \n2ha. The dotted line is an empirical lower limit
for SF galaxies (EW(8.6\mi) $>1\mu m$). 
{\it Bottom left:} \neii/\oiv\ versus \o3hb. SF galaxies are represented as black dots, composite galaxies as pink 
stars and AGNs as open red triangles. $r$ is the Pearson correlation coefficient.
{\it Right:} 8.6\mi\ PAH EWs against \neii/\oiv. 
This version of the \citet{Genzel_etal1998} diagram resembles a flipped version of the optical \cite{BPT1981} diagram.
The short-dashed lower line and the dot-dashed upper line are the Kewley et al. (2001) and Kauffmann et al. (2003) optical 
boundaries translated into the MIR plane using the correlations in the left panels.
The long-dashed middle line is an empiral boundary marking the region below which we do not find any optically defined SF galaxy.
\label{fig:fig2}
}
\end{figure}

\begin{figure}[t]
\plottwo{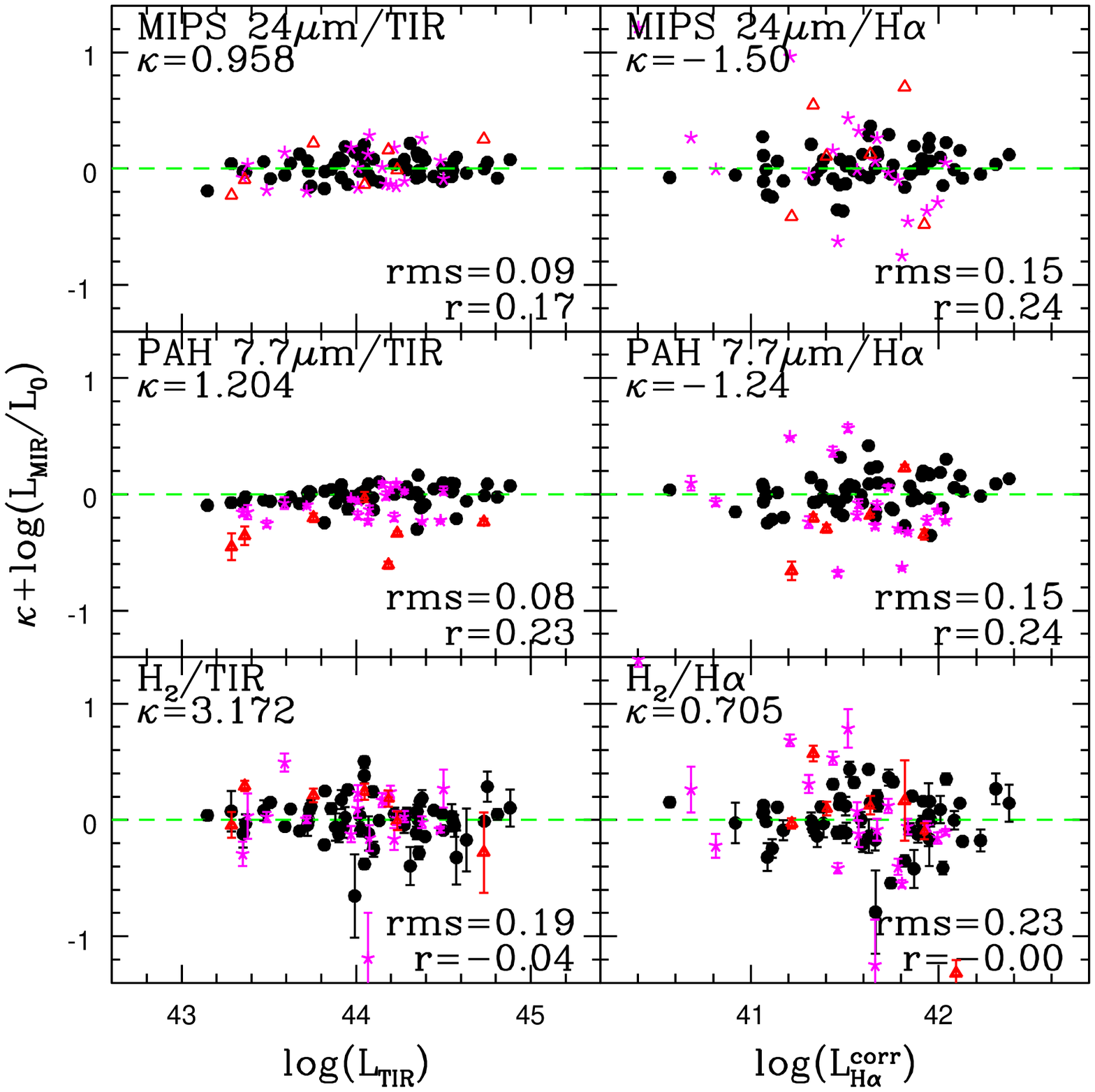}{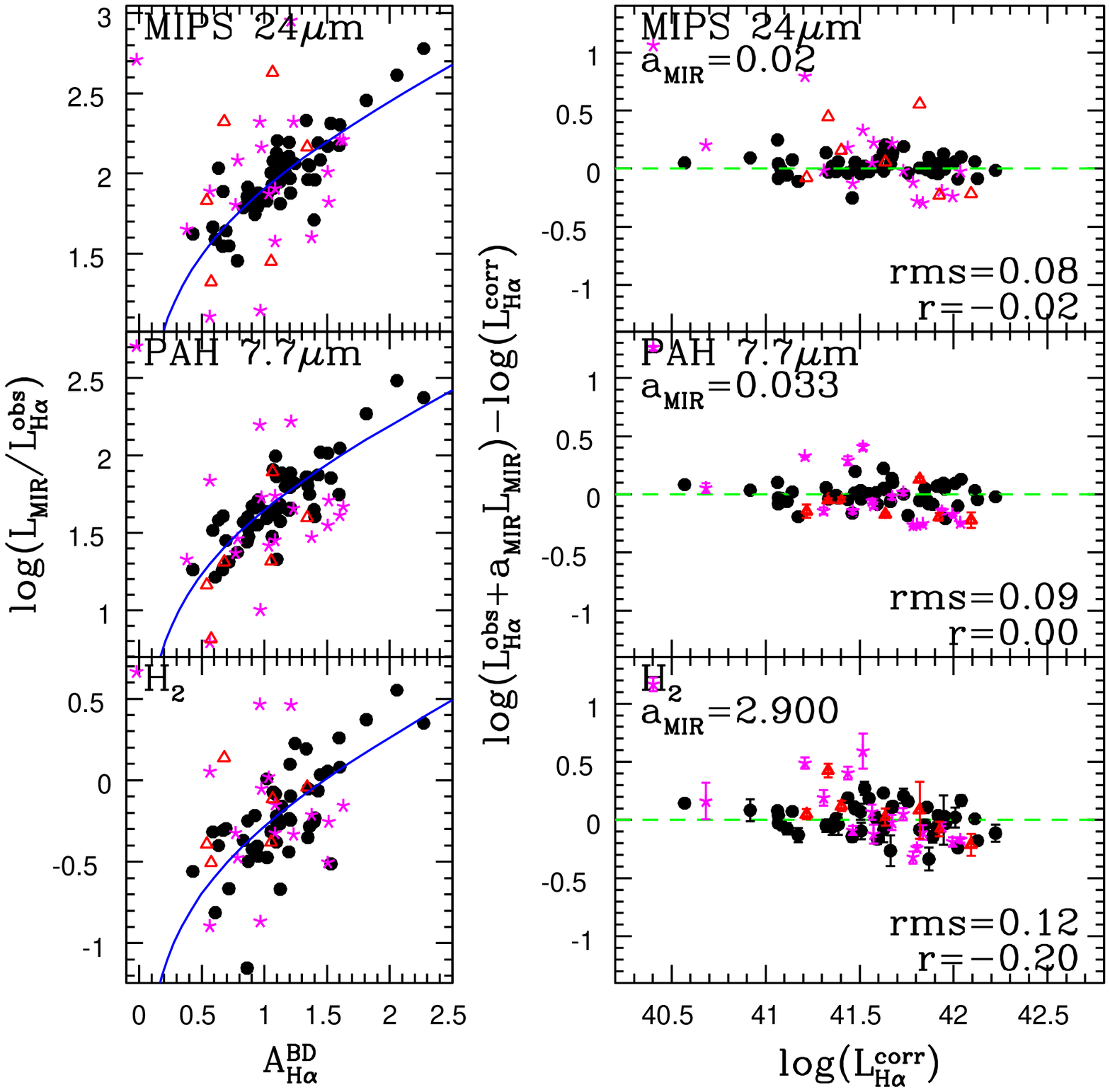}
\caption{{\it Column 1:} $L_{MIR}/L_{TIR}$ ratios as a function of \ltir\ where $L_{MIR}$ equals - from top to bottom - 
the 24\mi\ MIPS band luminosity ($\nu L_{\nu}$), the luminosity of the 7.7\mi\ PAH complex and the H$_2$ luminosity
where H$_2$ is defined as the sum of the $S(0)$ to $S(2)$ rotational lines.
The logarithmic scaling factors $\kappa$ are defined as the mean of log($L_{TIR}/L_{MIR}$) for the SF population alone 
(green dashed lines). The rms and Pearson coefficient $r$ are also for the SF population alone.
{\it Column 2:} $L_{MIR}/L_{H\alpha}^{corr}$ ratios as a function of $L_{H\alpha}^{corr}$ (the dust corrected \ha\ luminosity). 
{\it Column 3:} $L_{MIR}/L_{H\alpha}^{obs}$  ratios (observed \ha\ luminosity) against \ha\ attenuation measured from the Balmer decrement.
The solid lines are best fits to $A_{{\rm H}\alpha}=2.5~{\rm log}\left[1+ a_{MIR} {L_{MIR} / L_{{\rm H}\alpha}^{obs}}\right]$ \citep{Kennicutt_etal2009}.
{\it Column 4:} $L_{{\rm H}\alpha}^{obs}+a_{MIR} L_{MIR}$ to  $L_{{\rm H}\alpha}^{corr}$ ratios as a function of $L_{{\rm H}\alpha}^{corr}$. 
The $a_{MIR}$ coefficients derived in the previous column are indicated at the top left of each panel. 
\label{fig:fig3}
}
\end{figure}

\section{Conclusions}

$\bullet$ We find systematic trends between MIR spectral properties and optically derived physical properties, in particular 
between short wavelength PAH EWs and  \n2ha\ (gas phase metallicity), and between \neii/\oiv\ versus \o3hb\
(radiation field hardness) (Fig. 2, left panel); \\
$\bullet$ The Genzel et al. (1998) diagram has better resolution using PAHFIT than spline decompositions. It is very similar
to the optical ``BPT'' diagram (Fig. 2, right panel). The mixed SF/composite region may be revealing obscured AGNs 
in a large fraction of optically defined ``pure'' SF galaxies.\\
$\bullet$ The PAH, continuum, Ne and H$_2$ luminosities are tightly and nearly linearly correlated to the total IR luminosity, less so
to the dust corrected \ha\ luminosity (SFR) (Fig. 3, left panel);\\
$\bullet$ Following \cite{Kennicutt_etal2009}, the MIR components can be used to estimate dust attenuation in \ha\  and UV based on 
energy balance arguments (Fig. 3, right panel).




\end{document}